\pdfoutput=1
\documentclass[%
reprint,
superscriptaddress,
bibnotes,
longbibliography,
amsmath,amssymb,
pra,
floatfix,
]{revtex4-1}
\usepackage{graphicx}
\usepackage{dcolumn}
\usepackage{bm}
\usepackage{hyperref}

\usepackage{natbib}
\usepackage{mathrsfs,mathtools,color,wasysym,dsfont,here}
\usepackage{amsthm}
\usepackage{physics}
\usepackage[all]{xy}
\usepackage{tikz}
\usepackage{amsfonts}
\usepackage{array,booktabs}
\usepackage{comment}
\usetikzlibrary{calc}
\usetikzlibrary{arrows}
\usetikzlibrary{decorations.markings,decorations.pathmorphing}
\usepackage{cases}
\usepackage{siunitx}
\usepackage{tablefootnote}

\AtBeginDocument{
\heavyrulewidth=.08em
\lightrulewidth=.05em
\cmidrulewidth=.03em
\belowrulesep=.65ex
\belowbottomsep=0pt
\aboverulesep=.4ex
\abovetopsep=0pt
\cmidrulesep=\doublerulesep
\cmidrulekern=.5em
\defaultaddspace=.5em
}

\begin{document}

\title{Magnon-induced scalar spin chirality in Kagome and honeycomb ferromagnets}

\author{Nanse Esaki}
\email{esaki-nanse0428@g.ecc.u-tokyo.ac.jp}
\affiliation{Department of Physics, Graduate School of Science, The University of Tokyo, 7-3-1 Hongo, Tokyo 113-0033, Japan}
\affiliation{Advanced Science Research Center, Japan Atomic Energy Agency, Tokai, Ibaraki 319-1195, Japan}
\author{Gyungchoon Go}
\affiliation{Department of Physics, Korea Advanced Institute of Science and Technology, Daejeon 34141, Korea}
\author{Se Kwon Kim}
\email{sekwonkim@kaist.ac.kr}
\affiliation{Department of Physics, Korea Advanced Institute of Science and Technology, Daejeon 34141, Korea}

\begin{abstract}
The scalar spin chirality (SSC), defined as a triple product of spins, is essential for describing noncoplanar spin structures and understanding chiral physics in magnetic systems. Traditionally, SSC has been discussed primarily in the context of noncoplanar ground-state spin configurations at zero temperature, as collinear spin systems are generally thought to lack SSC. Consequently, whether the SSC can emerge at finite temperatures in spin systems with collinear ground states remains an open question and has yet to be fully understood. In this study, we theoretically demonstrate that thermally excited magnons can induce SSC even in collinear spin systems. By considering 2D ferromagnets on Kagome and honeycomb lattices, we demonstrate that the Dzyaloshinskii-Moriya interactions (DMI) which break the effective time-reversal symmetry in the magnon Hamiltonian can lead to finite SSC at finite temperatures. Using a simple spin model
, we show both numerically and analytically that the SSC increases with the magnitude of DMI and temperature. Furthermore, calculations based on realistic material parameters reveal that the magnon-induced SSC can achieve a magnitude comparable to those observed in non-coplanar spin configurations. These findings suggest that SSC plays a significant role even in collinear spin systems, providing new insights into the chiral physics of magnetic materials.
\end{abstract}

\maketitle

\section{Introduction}\label{sec:Intro}
The scalar spin chirality (SSC), defined as a triple product of spins $\chi_{ijk} = \bm{S}_i \cdot (\bm{S}_j \times \bm{S}_k)$, has emerged as a pivotal concept in the study of magnetic systems. SSC characterizes the fundamental unit of noncoplanar spin structures and is central to understanding the complex chiral physics inherent in magnetic materials. In the field of quantum magnetism, SSC has been introduced as a key descriptor of chiral spin states, which break time-reversal symmetry and are closely linked to unconventional phenomena such as chiral spin liquid and chiral superconductivity \cite{wen1989chiral}. Furthermore, SSC contributes to orbital magnetization via electron hopping among triplets of noncoplanar spins \cite{hoffmann2015topological, dos2016chirality, hanke2016role, lux2018engineering, grytsiuk2020topological, zhang2020imprinting}, providing a direct connection between spin chirality and electronic properties. SSC also plays a pivotal role in the transport properties of solid-state systems. It acts as an effective magnetic field for conduction electrons \cite{taguchi2001spin, tatara2002chirality, machida2010time} and magnons \cite{katsura2010theory}, inducing the Berry curvature in their band structures and driving Hall-like transport. Additionally, SSC gives rise to skew scattering of conduction electrons \cite{ishizuka2018spin} and phonons, which also results in Hall-type transport. These phenomena have been observed in various frustrated magnets \cite{taguchi2001spin, machida2010time, katsura2010theory, shindou2001orbital, nakatsuji2015large, han2017spin, kim2024thermal} and chiral magnets \cite{ishizuka2018spin, neubauer2009topological, kanazawa2011large, franz2014real, van2013magnetic, kim2019tunable, xu2024universal}, highlighting the critical importance of SSC across diverse material platforms. Consequently, understanding SSC offers a valuable perspective for exploring the rich chiral physics in both quantum and classical magnetic systems.

In ordered magnets, SSC is typically discussed within the framework of classical spin configurations, where spin systems with collinear ground states (which will be referred to as collinear spin systems below for simplicity) are generally thought to lack SSC. However, in collinear spin systems, Dzyaloshinskii-Moriya interactions (DMI) can break time-reversal symmetry in the magnon system, giving rise to the thermal Hall effect at finite temperatures \cite{katsura2010theory, onose2010observation, matsumoto2011theoretical, ideue2012effect, matsumoto2014thermal, mook2014magnon, mook2014edge, hirschberger2015thermal, kim2016realization, owerre2016first}. Given the close connection between SSC and Hall-like transport, it is reasonable to anticipate that thermally excited magnons might induce SSC even in collinear spin systems.
\begin{figure}[t]
    \centering
    \includegraphics[width=1.0\linewidth]{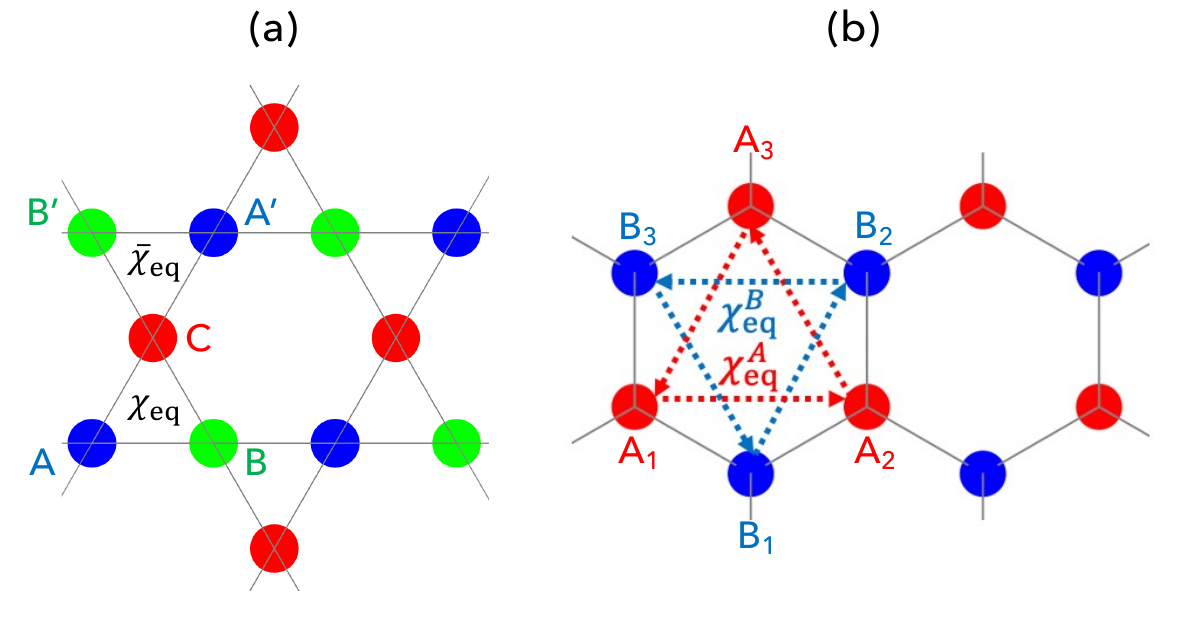}
    \caption{Schematic figure of Kagome (a) and Honeycomb (b) lattices. (a) SSC defined on the upper (lower) triangle is denoted as $\chi_{\mathrm{eq}}$ [Eq. (\ref{Eq:Def_chi})] ($\overline{\chi}_{\mathrm{eq}}$ [Eq. (\ref{Eq:Def_overline_chi})]). (b) SSC defined on the upper (lower) triangle is represented as $\chi^A_{\mathrm{eq}}$ [Eq. (\ref{Eq:Def_chi_A})] ($\chi^B_{\mathrm{eq}}$ [Eq. (\ref{Eq:Def_chi_B})]).}
    \label{fig:Model}
\end{figure}

In this work, we theoretically demonstrate that thermal magnons can induce SSC even in collinear spin systems by considering 2D ferromagnets on Kagome and honeycomb lattices [see Fig. \ref{fig:Model}]. These systems are widely recognized as canonical examples of systems exchibiting the thermal Hall effect of magnons. Using a simple spin model, we show that the DMI gives rise to a magnon-induced SSC at finite temperatures, even though the ground state remains collinear.
More specifically, in Kagome ferromagnets, the finite SSC emerges on the upper and lower triangles formed by the nearest-neighbor sites $ABC$ and $A'B'C$, respectively [see Fig. \ref{fig:Model} (a)], which is consistent with the previous report of the finite SSC in the paramagnetic phase studied by Monte Carlo calculations \cite{kolincio2023kagome}. In honeycomb ferromagnets, the finite SSC appears on the upper and lower triangles formed by the next-nearest-neighbor sites $A_1A_2A_3$ and $B_1B_2B_3$, respectively [see Fig. \ref{fig:Model} (b)]. Furthermore, we demonstrate both numerically and analytically that the SSC increases with the magnitude of DMI and temperature. Additionally, using calculations based on material parameters, we show that the magnon-induced SSC can reach a magnitude comparable to that of SSC observed in non-coplanar spin configurations in real materials at finite temperatures. This finding highlights that SSC can hold physical significance even in collinear spin systems.

The remainder of the paper is organized as follows. In Sec. \ref{subsec:Kagome_model}, we introduce the spin model used in this work and the SSC operators for Kagome ferromagnets. In Sec. \ref{subsec:Kagome_formalism}, we describe the analysis based on the Holstein-Primakoff transformation and present the formalism of the magnon-induced SSC, including its symmetry properties, in Kagome ferromagnets. In Sec. \ref{subsec:Kagome_results}, we provide the calculation results for the magnon band structure, the SSC profile, and the magnon-induced SSC in Kagome ferromagnets. From Sec. \ref{subsec:Honeycomb_model} to Sec. \ref{subsec:Honeycomb_results}, we extend our discussion to the magnon-induced SSC in honeycomb ferromagnets, following the same approach as in Kagome ferromagnets. Finally, in Sec. \ref{sec:Conclusion}, we conclude the paper by proposing possible experimental schemes and highlighting the important physical implications of our findings.

\section{Kagome ferromagnet case}\label{sec:Kagome}
\subsection{Model}\label{subsec:Kagome_model}
\indent Here we consider a 2D ferromagnet on a Kagome lattice [Fig. \ref{fig:Model} (a)] described by the spin Hamiltonian
\begin{equation}\label{Eq:Spin_Hamiltonian}
    \begin{split}
    \mathcal{H} &= -J\sum_{\langle {i,j} \rangle}\bm{S}_i \cdot \bm{S}_j - K \sum_{i} (S_{i,z})^2 + \sum_{\langle {i,j} \rangle}\bm{D}_{ij}\cdot(\bm{S}_i \times \bm{S}_j) \\ &- g\mu_B H \sum_{i} S_{i,z},
    \end{split}
\end{equation}
where $J (>0)$ is the ferromagnetic exchange coupling, $K (>0)$ is the easy-axis anisotropy, $\bm{D}_{ij} = D\hat{\bm{z}}$ is the DM vector pointing perpendicular to the plane, $g$ is the g-factor, $\mu_B$ is the Bohr magneton, and $H$ is the applied magnetic field along the $z$-direction. The sign convention for the DM vectors is chosen such that they are aligned along $+z (-z)$ for counterclockwise (clockwise) chirality: $A$-$B$-$C$ ($C$-$B$-$A$).

As shown in Fig. \ref{fig:Model} (a), the SSC operators $\chi$ and $\overline{\chi}$ are defined as the counterclockwise triple product of spins on the upper ($\triangle$) and lower ($\bigtriangledown$) triangles formed by the nearest-neighbor sites, i.e.,
\begin{equation}\label{Eq:Def_chi}
    \hat{\chi} = \frac{1}{N_{\mathrm{cell}}} \sum_{A, B, C \in \triangle} \frac{\bm{S}_A \cdot (\bm{S}_B \times \bm{S}_C)}{S^3},
\end{equation}
and
\begin{equation}\label{Eq:Def_overline_chi}
    \hat{\overline{\chi}} = \frac{1}{N_{\mathrm{cell}}} \sum_{A', B', C \in \bigtriangledown} \frac{\bm{S}_{A'} \cdot (\bm{S}_{B'} \times \bm{S}_C)}{S^3},
\end{equation}
respectively, where $N_{\mathrm{cell}}$ is the number of unit cells, and $S$ is the spin magnitude. Since the ground state of the system described by the Hamiltonian (\ref{Eq:Spin_Hamiltonian}) is collinear, \cite{mook2014magnon, mook2014edge} the SSC at zero temperature, i.e., the expectation value of Eq. (\ref{Eq:Def_chi}) or Eq. (\ref{Eq:Def_overline_chi}) in the ground state vanishes.

\subsection{Formalism}\label{subsec:Kagome_formalism}
At low temperatures, the thermodynamic properties of ferromagnets are well described by low-energy excitations known as magnons. Thus, it is natural to formulate the SSC at finite temperatures based on the magnon eigenstates \cite{go2024scalar}. To describe such magnon excitations, we employ the Holstein-Primakoff transformation, expressed as $S^{+}_i = \sqrt{2S-a^{\dag}_i a_i}a_i$, $S^{-}_i = a^{\dag}_i \sqrt{2S-a^{\dag}_i a_i}$, and $S^{z}_i = S- a^{\dag}_i a_i$, where $a_i$ and $a^{\dag}_i$ are the annihilation and creation operators of a magnon at the site $i$. By introducing the Fourier transformation $a_i = \sum_{\bm{k}} e^{i\bm{k}\cdot \bm{R}_i} a_{\alpha\bm{k}}/\sqrt{N_{\mathrm{cell}}}$ ($i \in \alpha, \quad \alpha = \mathcal{A}, \mathcal{B}, \mathcal{C}$), where $\mathcal{A}$, $\mathcal{B}$, and $\mathcal{C}$ represent the sublattice indices, and retaining only terms up to quadratic order in the magnon operators, we obtain:

\begin{equation}\label{Eq:Magnon_Hamiltonian_Kagome}
    \mathcal{H} = \sum_{\bm{k}} \Psi^{\dag}_{\bm{k}} h_{\bm{k}} \Psi_{\bm{k}}, \quad \Psi_{\bm{k}} = (a_{\mathcal{A}\bm{k}},a_{\mathcal{B}\bm{k}},a_{\mathcal{C}\bm{k}})^{T},
\end{equation}
with the momentum space Hamiltonian

\begin{equation}\label{Eq:hk_Kagome}
    h_{\bm{k}} = \overline{\epsilon} - 2S\sqrt{J^2 + D^2} \begin{pmatrix}
        0 & \cos{k_1} e^{i\phi} & \cos{k_2} e^{-i\phi}\\
        \cos{k_1}e^{-i\phi} & 0 & \cos{k_3}e^{i\phi}\\
        \cos{k_2}e^{i\phi} & \cos{k_3}e^{-i\phi} & 0
    \end{pmatrix},
\end{equation}
where $\overline{\epsilon} = 2S(2J+K)+g\mu_B H$, $\phi = \tan^{-1}(D/J)$, and $k_j = \bm{k} \cdot \bm{a}_j$ ($j = 1, 2, 3$) with $\bm{a}_1 = (1,0)$, $\bm{a}_2 = \frac{1}{2}(1,\sqrt{3})$, and $\bm{a}_3 = \frac{1}{2}(1,-\sqrt{3})$. By diagonalizing the above Hamiltonian (\ref{Eq:hk_Kagome}), we obtain its eigenvalues and eigenstates $\epsilon_{n,\bm{k}}$ and $\ket{u_{n,\bm{k}}}$ ($\epsilon_{1,\bm{k}} \leq \epsilon_{2,\bm{k}} \leq \epsilon_{3,\bm{k}}$), respectively.

In terms of the magnon operators, the SSC operators (\ref{Eq:Def_chi}) and (\ref{Eq:Def_overline_chi}) are expressed in the following form:

\begin{equation}\label{Eq:Magnon_SSC_Kagome}
  \hat{\chi} = \frac{1}{N_{\mathrm{cell}}} \sum_{\bm{k}} \Psi^{\dag}_{\bm{k}} \hat{\chi}_{\bm{k}} \Psi_{\bm{k}}, \quad \Psi_{\bm{k}} = (a_{\mathcal{A}\bm{k}},a_{\mathcal{B}\bm{k}},a_{\mathcal{C}\bm{k}})^{T},
\end{equation}
and
\begin{equation}\label{Eq:Magnon_overline_SSC_Kagome}
  \hat{\overline{\chi}} = \frac{1}{N_{\mathrm{cell}}} \sum_{\bm{k}} \Psi^{\dag}_{\bm{k}} \hat{\overline{\chi}}_{\bm{k}} \Psi_{\bm{k}}, \quad \Psi_{\bm{k}} = (a_{\mathcal{A}\bm{k}},a_{\mathcal{B}\bm{k}},a_{\mathcal{C}\bm{k}})^{T},
\end{equation}
with the SSC matrices
\begin{equation}\label{Eq:chi_k_Kagome}
    \hat{\chi}_{\bm{k}} = \frac{-i}{S} \begin{pmatrix}
        0 & e^{i k_1} & -e^{i k_2}\\ 
        -e^{-i k_1} & 0 & e^{-i k_3}\\
        e^{-i k_2} & -e^{i k_3} & 0
    \end{pmatrix},
\end{equation}
and
\begin{equation}\label{Eq:overline_chi_k_Kagome}
    \begin{split}
    \hat{\overline{\chi}}_{\bm{k}} &= \frac{-i}{S} \begin{pmatrix}
        0 & e^{-i k_1} & -e^{-i k_2}\\ 
        -e^{i k_1} & 0 & e^{i k_3}\\
        e^{i k_2} & -e^{-i k_3} & 0
    \end{pmatrix} \\
    &= \hat{\chi}_{-\bm{k}},
    \end{split}
\end{equation}
respectively. 

Using the above expressions (\ref{Eq:chi_k_Kagome}) and (\ref{Eq:overline_chi_k_Kagome}), we formulate the magnon-induced SSC at finite temperatures by
\begin{equation}\label{Eq:Magnon_induced_SSC_def_Kagome}
    \chi_{\mathrm{eq}} = \frac{1}{N_{\mathrm{cell}}} \sum_{\bm{k}} \rho (\epsilon_{n,\bm{k}}) \chi_{n,\bm{k}},
\end{equation}
and
\begin{equation}\label{Eq:Magnon_induced_overline_SSC_def_Kagome}
    \overline{\chi}_{\mathrm{eq}} = \frac{1}{N_{\mathrm{cell}}} \sum_{\bm{k}} \rho (\epsilon_{n,\bm{k}}) \overline{\chi}_{n,\bm{k}},
\end{equation}
where $\chi_{n,\bm{k}} = \mel{u_{n,\bm{k}}}{\hat{\chi}_{\bm{k}}}{u_{n,{\bm{k}}}}$, $\overline{\chi}_{n,\bm{k}} = \mel{u_{n,\bm{k}}}{\hat{\overline{\chi}}_{\bm{k}}}{u_{n,{\bm{k}}}}$, and $\rho(\epsilon_{n,\bm{k}}) = 1/(e^{\beta \epsilon_{n,\bm{k}}}-1)$ is the Bose distribution function with $\beta$ being the inverse temperature. Here we note that $\chi_{n,\bm{k}}$ coincides with $\overline{\chi}_{n,-\bm{k}}$ due to the relations $\ket{u_{n,\bm{k}}} = \ket{u_{n,-\bm{k}}}$ and $\hat{\chi}_{\bm{k}} = \hat{\overline{\chi}}_{-\bm{k}}$. This reflects the fact that the upper and lower triangles in Fig. \ref{fig:Model} (a) are connected by spatial inversion. As a result, $\chi_{\mathrm{eq}} = \overline{\chi}_{\mathrm{eq}}$ and in the following discussion, we shall focus exclusively on $\chi_{n,\bm{k}}$ and $\chi_{\mathrm{eq}}$.

Here, we discuss the symmetry properties of $\chi_{\mathrm{eq}}$. Under the inversion operation $h_{\bm{k}} \rightarrow h_{-\bm{k}}$, $\epsilon_{n, \bm{k}}$ and $\chi_{n,\bm{k}}$ are transformed as $\epsilon_{n,-\bm{k}}$ and $\chi_{n,-\bm{k}}$, respectively. This suggests that the SSC (\ref{Eq:Magnon_induced_SSC_def_Kagome}) is invariant under the inversion operation. On the other hand, under the time-reversal operation $h_{\bm{k}} \rightarrow h^{\ast}_{-\bm{k}}$, $\epsilon_{n, \bm{k}}$ and $\chi_{n,\bm{k}}$ are transformed as $\epsilon_{n,-\bm{k}}$ and $\mel{u^{\ast}_{n,-\bm{k}}}{\hat{\chi}_{\bm{k}}}{u^{\ast}_{n,{-\bm{k}}}}$, respectively. By applying the relation $\hat{\chi}_{\bm{k}} = -\hat{\chi}^{\ast}_{-\bm{k}}$, we find that $\chi_{\mathrm{eq}}$ transforms to $-\chi_{\mathrm{eq}}$ under the time-reversal operation, suggesting that the SSC vanishes if $h_{\bm{k}}$ preserves the time-reversal symmetry. Therefore, the DMI, which breaks the time-reversal symmetry for the magnon Hamiltonian (\ref{Eq:hk_Kagome}) is necessary for the finite SSC at finite temperatures.

\subsection{Results}\label{subsec:Kagome_results}
\begin{figure}[t]
    \centering
    \includegraphics[width=1.0\linewidth]{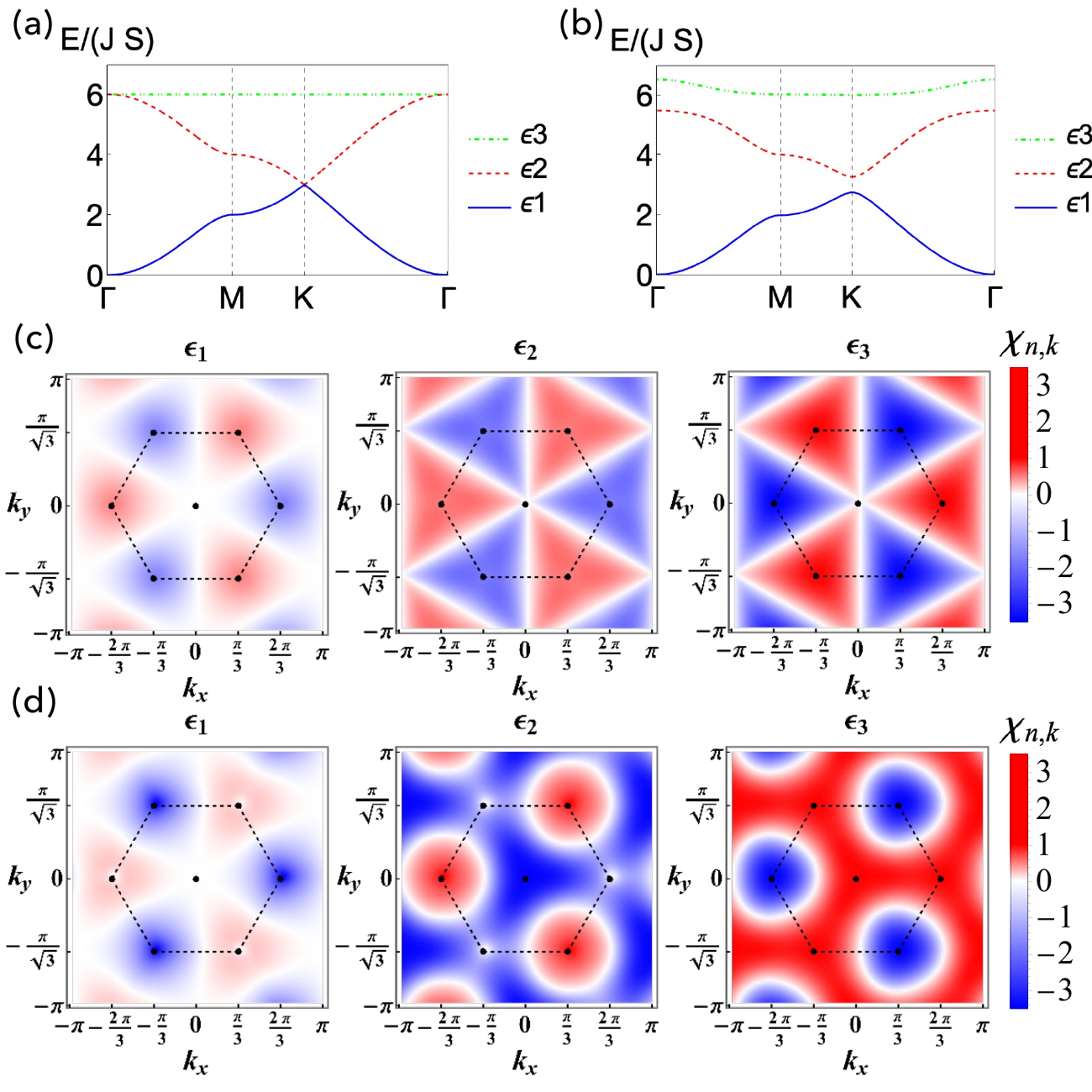}
    \caption{Magnon band structure and SSC profile in the case of Kagome ferromagnets. (a, b) Band structure (a) without DMI and (b) with DMI ($D_z = \SI{0.09}{meV}$). (c, d) Equilibrium SSC profile in momentum space $\chi_{n,\bm{k}}$ ($n = 1, 2, 3$) (c) without DMI and (d) with DMI ($D_z = \SI{0.09}{meV}$). The parameters used in the calculations are $S = 1/2$, $J = \SI{0.6}{meV}$, $K = 0$, and $H = \SI{0.05}{T}$.}
    \label{fig:Band_texture_Kagome}
\end{figure}

Figure \ref{fig:Band_texture_Kagome} displays the magnon band structure and the SSC profile $\chi_{n,\bm{k}}$ both without and with DMI. In Fig. \ref{fig:Band_texture_Kagome}, the high-symmetry points in reciprocal space are denoted by $\Gamma = (0,0)$, $M = (\frac{\pi}{2}, \frac{\pi}{2\sqrt{3}})$, $K = (\frac{\pi}{3}, \frac{\pi}{\sqrt{3}})$. For the numerical calculations, we use the parameters of Cu(1,3-bdc): $S = 1/2$, $J = \SI{0.6}{meV}$, $K = 0$ \cite{hirschberger2015thermal, chisnell2015topological}. In Fig. \ref{fig:Band_texture_Kagome} (b), we see that $\chi_{n,\bm{k}}$ is an odd function with respect to $\bm{k}$ with $C_3$ rotational symmetry. This observation is consistent with the symmetry argument in Sec. \ref{subsec:Kagome_formalism} and the fact that the SSC operator (\ref{Eq:Def_chi}) remains invariant under cyclic permutations \cite{go2024scalar}. On the other hand, Fig. \ref{fig:Band_texture_Kagome} (d) shows that $\chi_{n,\bm{k}}$ is no longer an odd function, reflecting the breaking of time-reversal symmetry by the DMI in the magnon Hamiltonian (\ref{Eq:hk_Kagome}).

\begin{figure}[H]
    \centering
    \includegraphics[width=1.0\linewidth]{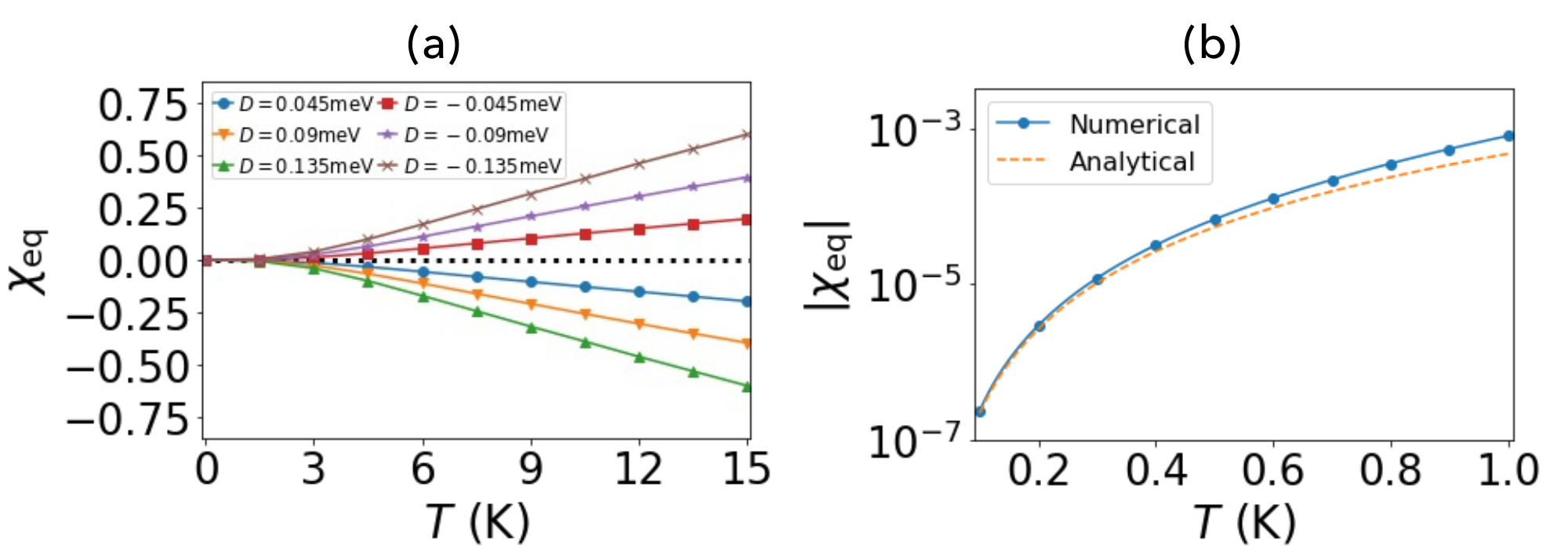}
    \caption{Temperature and DMI dependence of the magnon-induced SSC in Kagome ferromagnets (\ref{Eq:Magnon_induced_SSC_def_Kagome}). (a) Numerical results for varying DMI. (b) Comparison between the numerical results and the analytical expression (\ref{Eq:SSC_eq_analytical_Kagome}) with $D_z = \SI{0.09}{meV}$ below $\SI{1}{K}$. The parameters used in the calculations are $S = 1/2$, $J = \SI{0.6}{meV}$, $K = 0$, and $H = \SI{0.05}{T}$.}
    \label{fig:SSC_Numerical_Kagome}
\end{figure}

Figure \ref{fig:SSC_Numerical_Kagome} shows the temperature and DMI dependence of $\chi_{\mathrm{eq}}$ [Eq. (\ref{Eq:Magnon_induced_SSC_def_Kagome})] calculated using the parameters of Cu(1,3-bdc). Figure \ref{fig:SSC_Numerical_Kagome} (a) indicates that $\abs{\chi_{\mathrm{eq}}}$ increases with increasing temperatures, and $\chi_{\mathrm{eq}}$ depends almost linearly on the DMI. To explain this behavior, we derive the following analytical expression in the low-temperature limit $k_B T \lessapprox 2KS+g\mu_B H $: 
\begin{equation}\label{Eq:SSC_eq_analytical_Kagome}
\chi_{\mathrm{eq}} \simeq -\frac{\sqrt{3} D_z}{9\pi\sqrt{J^2 + D^2_z}S} \frac{e^{-\beta(2KS + g\mu_B H)}}{(\beta JS)^3}.
\end{equation}
See Appendix \ref{Appendix:Derivation_Kagome} for detailed derivation. As shown in Fig. \ref{fig:SSC_Numerical_Kagome} (b), this expression (\ref{Eq:SSC_eq_analytical_Kagome}) is consistent with the numerical results at sufficiently low temperatures. 

To discuss the experimental feasibility of the magnon-induced SSC, we compare our results shown in Fig. \ref{fig:SSC_Numerical_Kagome} with the SSC in Kagome antiferromagnets with non-coplanar spin configurations, which is expressed as \cite{laurell2018magnon}
\begin{equation} 
\chi_{\mathrm{KAFM}} = \frac{3\sqrt{3}}{2} \cos^{2}(\eta)\sin(\eta), 
\end{equation} 
where $\eta$ is the canting angle. In some realistic Kagome antiferromagnets with non-coplanar spin configurations identified in experiments, canting angles are on the order of $\ang{1}$ \cite{laurell2018magnon}, corresponding to $\chi_{\mathrm{KAFM}}\sim 0.045$. Therefore, the magnon-induced SSC in Kagome ferromagnets, shown in Fig. \ref{fig:SSC_Numerical_Kagome}, is comparable to the SSC in realistic Kagome antiferromagnets. This is one of our main results: Kagome ferromagnets with collinear ground states possess the finite SSC induced by thermal magnons at finite temperatures and the magnitude of SSC can be comparable to that of Kagome antiferromagnets. This point is discussed further in Sec. \ref{sec:Conclusion}.


\section{Honeycomb ferromagnet case}
\subsection{Model}\label{subsec:Honeycomb_model}
\indent Here we consider a 2D ferromagnet on a honeycomb lattice [Fig. \ref{fig:Model} (b)] described by the spin Hamiltonian (\ref{Eq:Spin_Hamiltonian}). As in the case of Kagome ferromagnets, the DM vectors are defined such that their alignment is along $+z (-z)$ for counterclockwise (clockwise) chirality, corresponding to $A_1$-$A_2$-$A_3$ and $B_1$-$B_2$-$B_3$ ($A_3$-$A_2$-$A_1$ and $B_3$-$B_2$-$B_1$).

As shown in Fig. \ref{fig:Model} (b), the SSC operators $\chi^A$ and $\chi^B$ are defined as the counterclockwise triple product of spins on the upper ($\triangle$) and lower ($\bigtriangledown$) triangles formed by the next-nearest-neighbor sites, i.e.,
\begin{equation}\label{Eq:Def_chi_A}
    \hat{\chi}^A =  \frac{1}{N_{\mathrm{cell}}} \sum_{A_1, A_2, A_3 \in \triangle} \frac{\bm{S}_{A_1} \cdot (\bm{S}_{A_2} \times \bm{S}_{A_3})}{S^3},
\end{equation}
and
\begin{equation}\label{Eq:Def_chi_B}
    \hat{\chi}^B =  \frac{1}{N_{\mathrm{cell}}} \sum_{B_1, B_2, B_3 \in \bigtriangledown} \frac{\bm{S}_{B_1} \cdot (\bm{S}_{B_2} \times \bm{S}_{B_3})}{S^3},
\end{equation}
respectively. As mentioned in Sec. \ref{subsec:Kagome_model}, the SSC is also absent in honeycomb ferromagnets at zero temperature.

Here, we note that SSC is typically defined using nearest-neighbor spins, whereas in honeycomb ferromagnets, it is defined using next-nearest-neighbor spins. This definition is reasonable, given that SSC acts as an effective magnetic field for hopping particles. In honeycomb ferromagnets, SSC defined by next-nearest-neighbor spins indeed functions as an effective magnetic field, similar to the Haldane model, a prototypical example of a topological insulator \cite{haldane1988model, kim2016realization, owerre2016first}.

\subsection{Formalism}\label{subsec:Honeycomb_formalism}
Applying the same procedure as in Sec. \ref{subsec:Kagome_formalism}, the spin Hamiltonian (\ref{Eq:Spin_Hamiltonian}) can be written in terms of the magnon operators as

\begin{equation}\label{Eq:Magnon_Hamiltonian_Honeycomb}
    \mathcal{H} = \sum_{\bm{k}} \Psi^{\dag}_{\bm{k}} h_{\bm{k}} \Psi_{\bm{k}}, \quad \Psi_{\bm{k}} = (a_{A\bm{k}},a_{B\bm{k}})^{T},
\end{equation}
with the momentum space Hamiltonian

\begin{equation}\label{Eq:hk_Honeycomb}
    h_{\bm{k}} = \tilde{\epsilon} - S \sum_{j=1}^{3} \begin{pmatrix}
        2D_z \sin(\bm{k}\cdot \bm{d}_j) & J\exp(i\bm{k}\cdot \bm{c}_j)\\
        J\exp(-i\bm{k}\cdot \bm{c}_j) & 2D_z \sin(-\bm{k}\cdot \bm{d}_j)\\
    \end{pmatrix},
\end{equation}
where $A$ and $B$ denote the sublattice indices, $\tilde{\epsilon} = S(3J+2K)+g\mu_B H$, $\bm{c}_1 = (0,1)$, $\bm{c}_2 = \frac{1}{2}(-\sqrt{3},-1)$, $\bm{c}_3 = \frac{1}{2}(\sqrt{3},-1)$, $\bm{d}_1 = (-\sqrt{3},0)$, $\bm{d}_2 = \frac{1}{2}(\sqrt{3},-3)$, and $\bm{d}_3 = \frac{1}{2}(\sqrt{3},3)$. Diagonalizing the above Hamiltonian (\ref{Eq:hk_Honeycomb}) yields its eigenvalues and eigenstates $\epsilon_{n,\bm{k}}$ and $\ket{u_{n,\bm{k}}}$ ($\epsilon_{1,\bm{k}} \leq \epsilon_{2,\bm{k}}$), respectively.

In terms of the magnon operators, the SSC operators (\ref{Eq:Def_chi_A}) and (\ref{Eq:Def_chi_B}) take the following form:

\begin{equation}\label{Eq:Magnon_SSC_Honeycomb}
  \chi^{A/B} = \frac{1}{N_{\mathrm{cell}}} \sum_{\bm{k}} \Psi^{\dag}_{\bm{k}} \hat{\chi}^{A/B}_{\bm{k}} \Psi_{\bm{k}}, \quad \Psi_{\bm{k}} = (a_{A\bm{k}},a_{B\bm{k}})^{T},
\end{equation}
with the SSC matrices
\begin{align}\label{Eq:chi_k_Honeycomb_A}
    \hat{\chi}^{A}_{\bm{k}} &= \sum_{j=1}^{3} \frac{2}{S} \begin{pmatrix}
        \sin(\bm{k}\cdot \bm{d}_j) & 0 \\
        0 & 0 \\
    \end{pmatrix},\\ \label{Eq:chi_k_Honeycomb_B}
    \hat{\chi}^{B}_{\bm{k}} &= -\sum_{j=1}^{3} \frac{2}{S} \begin{pmatrix}
        0 & 0 \\
        0 & \sin(\bm{k}\cdot \bm{d}_j) \\
    \end{pmatrix},\\ \notag
\end{align}
respectively.

From the above expressions (\ref{Eq:chi_k_Honeycomb_A}) and (\ref{Eq:chi_k_Honeycomb_B}), we formulate the magnon-induced SSC at finite temperatures for each sublattice by

\begin{equation}\label{Eq:Magnon_induced_SSC_A_def_Honeycomb}
    \chi^{A}_{\mathrm{eq}} = \frac{1}{N_{\mathrm{cell}}} \sum_{\bm{k}} \rho (\epsilon_{n,\bm{k}}) \chi^{A}_{n,\bm{k}},
\end{equation}
and
\begin{equation}\label{Eq:Magnon_induced_SSC_B_def_Honeycomb}
    \chi^{B}_{\mathrm{eq}} = \frac{1}{N_{\mathrm{cell}}} \sum_{\bm{k}} \rho (\epsilon_{n,\bm{k}}) \chi^{B}_{n,\bm{k}},
\end{equation}
respectively, where $\chi^{A}_{n,\bm{k}} = \mel{u_{n,\bm{k}}}{\hat{\chi}^{A}_{\bm{k}}}{u_{n,{\bm{k}}}}$ and $\chi^{B}_{n,\bm{k}} = \mel{u_{n,\bm{k}}}{\hat{\chi}^{B}_{\bm{k}}}{u_{n,{\bm{k}}}}$. Hereafter, we shall focus only on $\chi^{A}_{\mathrm{eq}}$ since $\chi^{A}_{\mathrm{eq}}$ takes the same value as that of $\chi^{B}_{\mathrm{eq}}$. This can be understood as follows: Under the inversion operation $\bm{k}\rightarrow -\bm{k}$ and the exchange of sublattice indices $A\leftrightarrow B$, the Hamiltonian (\ref{Eq:hk_Honeycomb}) remains invariant. However, the SSC matrices (\ref{Eq:chi_k_Honeycomb_A}) and (\ref{Eq:chi_k_Honeycomb_B}) are interchanged as $\hat{\chi}^{A}_{\bm{k}} \leftrightarrow \hat{\chi}^{B}_{\bm{k}}$. This implies that $\chi^{A}_{n,\bm{k}} = \chi^{B}_{n,\bm{k}}$, leading to $\chi^{A}_{\mathrm{eq}} = \chi^{B}_{\mathrm{eq}}$.
\begin{figure}[t]
    \centering
    \includegraphics[width=0.9\linewidth]{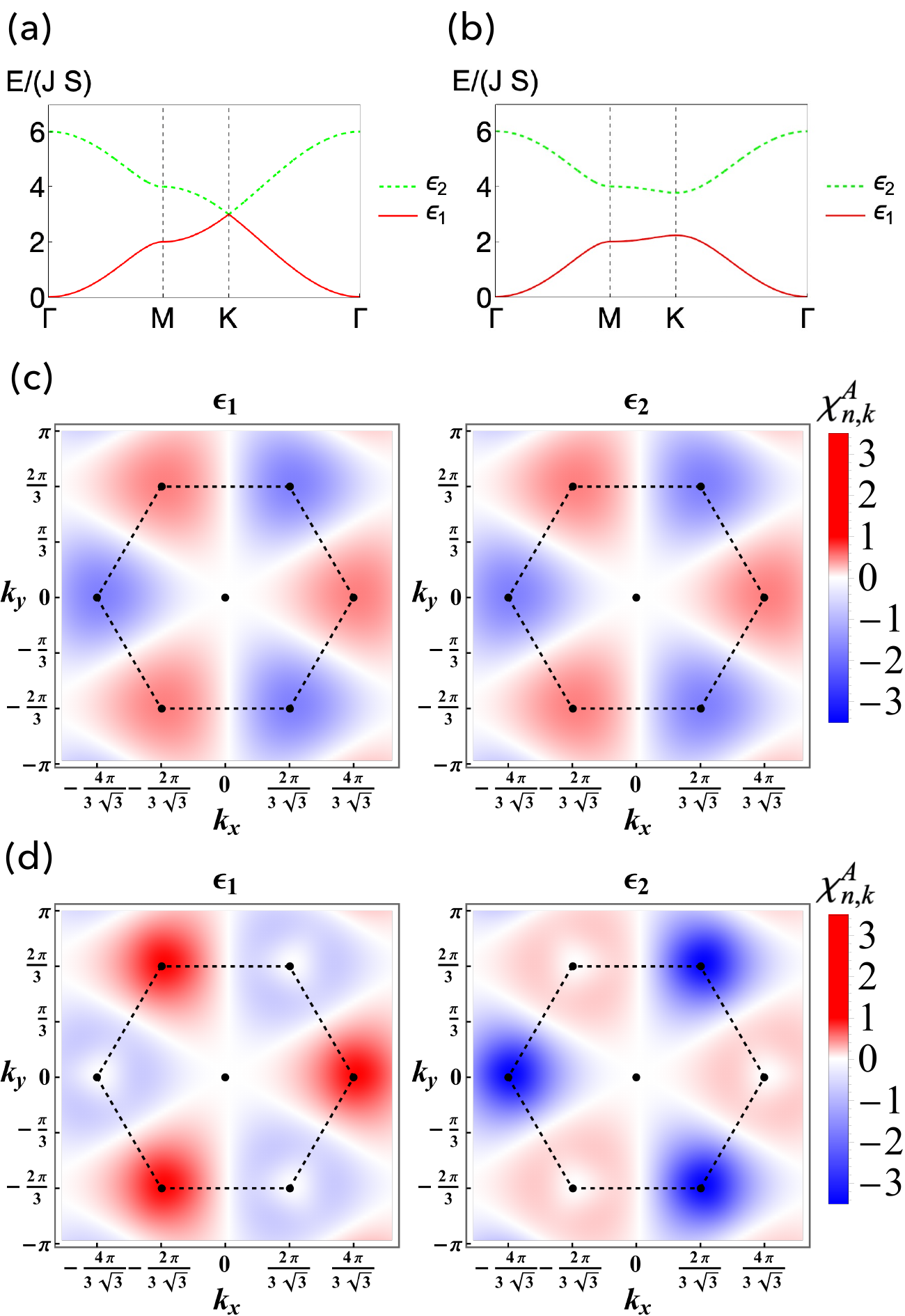}
    \caption{Magnon band structure and SSC profile in the case of honeycomb ferromagnets. (a, b) Band structure (a) without DMI and (b) with DMI ($D_z = \SI{0.22}{meV}$). (c, d) Equilibrium SSC profile in momentum space $\chi^{A}_{n,\bm{k}}$ ($n = 1, 2$) (c) without DMI and (d) with DMI ($D_z = \SI{0.22}{meV}$). The parameters used in the calculations are $S = 3/2$, $J = \SI{1.48}{meV}$, $K = \SI{0.02}{meV}$, and $H = 0$.}
    \label{fig:Band_texture_Honeycomb}
\end{figure}

It should be noted that since the SSC matrix (\ref{Eq:chi_k_Honeycomb_A}) satisfies $\chi^{A}_{\bm{k}} = -\chi^{A*}_{-\bm{k}}$, the time-reversal symmetry of the magnon Hamiltonian (\ref{Eq:hk_Honeycomb}) results in the vanishing SSC (\ref{Eq:Magnon_induced_SSC_A_def_Honeycomb}), as discussed in Sec. \ref{subsec:Kagome_formalism}. Consequently, the DMI, which breaks the time-reversal symmetry of the magnon Hamiltonian (\ref{Eq:hk_Honeycomb}), plays a crucial role in inducing the finite SSC also in the case of honeycomb ferromagnets.

\subsection{Results}\label{subsec:Honeycomb_results}
Figure \ref{fig:Band_texture_Honeycomb} shows the magnon band structure and the SSC profile $\chi^{A}_{n,\bm{k}}$ with and without DMI, using the parameters of CrBr$_3$: $S = 3/2$, $J = \SI{1.48}{meV}$, $K = \SI{0.02}{meV}$ \cite{cai2021topological}. In Fig. \ref{fig:Band_texture_Honeycomb}, the high-symmetry points in reciprocal space are represented by $\Gamma = (0,0)$, $M = (\frac{\pi}{\sqrt{3}}, \frac{\pi}{3})$, $K = (\frac{2\pi}{3\sqrt{3}}, \frac{2\pi}{3})$. In the absence of DMI, as shown in Fig. \ref{fig:Band_texture_Honeycomb} (b), $\chi^{A}_{n,\bm{k}}$ is an odd function of $\bm{k}$ with $C_3$ rotational symmetry. With DMI, however, $\chi^{A}_{n,\bm{k}}$ loses this odd symmetry, as seen in Fig. \ref{fig:Band_texture_Honeycomb} (d), indicating the violation of time-reversal symmetry caused by the DMI in the magnon Hamiltonian (\ref{Eq:hk_Honeycomb}). This behavior aligns with the results for Kagome ferromagnets shown in Fig. \ref{fig:Band_texture_Kagome}.

\begin{figure}[t]
    \centering
    \includegraphics[width=1.0\linewidth]{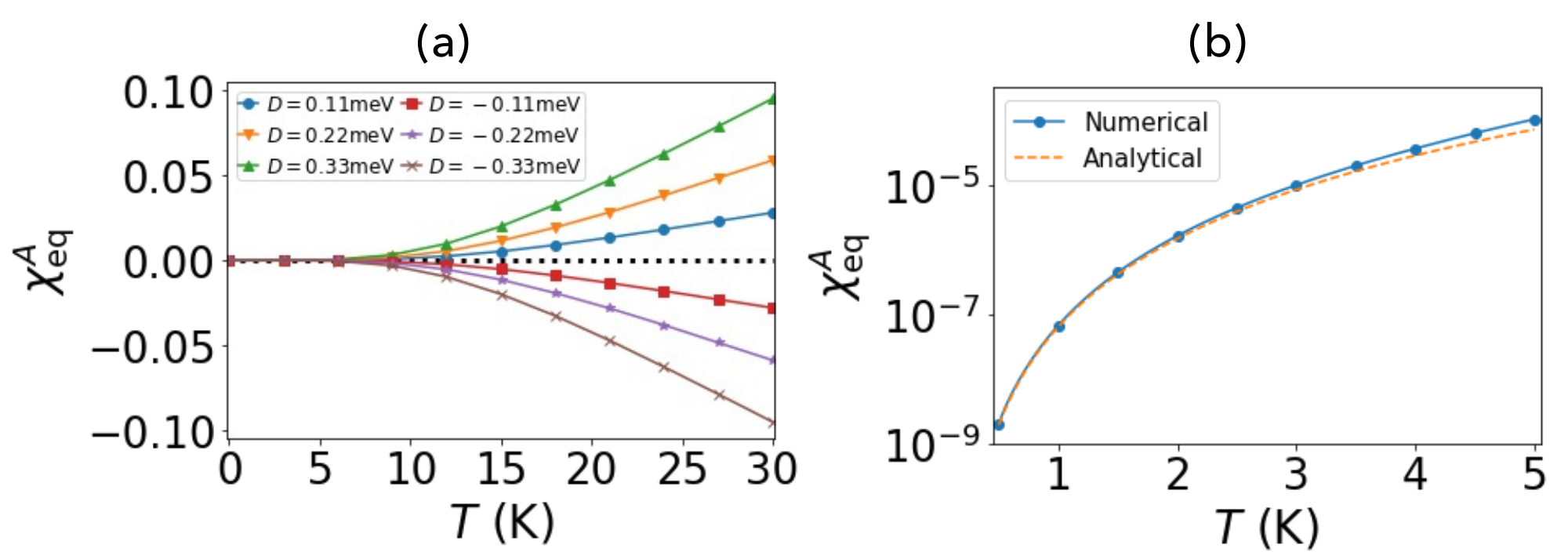}
    \caption{Temperature and DMI dependence of the magnon-induced SSC in honeycomb ferromagnets (\ref{Eq:Magnon_induced_SSC_A_def_Honeycomb}). (a) Numerical results for varying DMI. (b) Comparison between the numerical results and the analytical expression (\ref{Eq:SSC_eq_analytical_Honeycomb}) with $D_z = \SI{0.22}{meV}$ below $\SI{5}{K}$. The parameters used in the calculations are $S = 3/2$, $J = \SI{1.48}{meV}$, $K = \SI{0.02}{meV}$, and $H = 0$.}
    \label{fig:SSC_Numerical_Honeycomb}
\end{figure}

Figure \ref{fig:SSC_Numerical_Honeycomb} illustrates the temperature and DMI dependence of $\chi^{A}_{\mathrm{eq}}$, calculated using the parameters of CrBr$_3$. Fig. \ref{fig:SSC_Numerical_Honeycomb} (a) suggests that $\chi^{A}_{\mathrm{eq}}$ exhibits a similar dependence on temperature and DMI as in the case of Kagome ferromagnets [see also Fig. \ref{fig:SSC_Numerical_Kagome}]. This similarity can be understood from the following analytical expression, which is derived in the same way as in the derivation of Eq. (\ref{Eq:SSC_eq_analytical_Kagome}):
\begin{equation}\label{Eq:SSC_eq_analytical_Honeycomb}
\chi^{A}_{\mathrm{eq}} \simeq \frac{\sqrt{3} D_z}{\pi JS} \frac{e^{-\beta(2KS+g\mu_B H)}}{(\beta JS)^4}.
\end{equation}
See Appendix \ref{Appendix:Derivation_Honeycomb} for detailed derivation. This expression (\ref{Eq:SSC_eq_analytical_Honeycomb}) agrees well with the numerical calculations, as demonstrated in Fig. \ref{fig:SSC_Numerical_Honeycomb} (b). The main difference from the case of Kagome ferromagnets lies in the power-law temperature dependence: while the SSC in Kagome ferromagnets (\ref{Eq:SSC_eq_analytical_Kagome}) is proportional to $T^3$, that in honeycomb ferromagnets (\ref{Eq:SSC_eq_analytical_Honeycomb}) is proportional to $T^4$. Notably, the magnitude of the magnon-induced SSC in a honeycomb ferromagnet is also on the order of $0.1$, indicating that honeycomb ferromagnets are likewise promising candidates for observing magnon-induced SSC.

\section{Conclusion and outlook}\label{sec:Conclusion}
In this work, we have demonstrated that the magnon-induced SSC can emerge in Kagome and honeycomb ferromagnets with the DMI. In the absence of the DMI, the time-reversal symmetry of the magnon Hamiltonian makes the SSC profile in momentum space antisymmetric with respect to the crystal momentum, leading to the absence of the SSC in both models. However, the finite DMI breaks this symmetry, inducing the SSC at finite temperatures even in ferromagnets with a collinear ground state. We have demonstrated, both numerically and analytically, that such magnon-induced SSC develops with the magnitude of DMI and temperature. Furthermore, we have numerically shown that the magnon-induced SSC in Kagome and honeycomb ferromagnets with realistic parameters can be comparable to SSC observed in realistic Kagome antiferromagnets, suggesting its experimental feasibility.

Therefore, we expect that the magnon-induced SSC can be observed in Kagome ferromagnets like Cu(1,3-bdc) \cite{hirschberger2015thermal, chisnell2015topological} and honeycomb ferromagnets including Cr$X_3$ ($X = \mathrm{Br}, \mathrm{I}$) \cite{chen2018topological, kvashnin2020relativistic, chen2021magnetic, cai2021topological} and Cr$X$Te$_3$ ($X = \mathrm{Ge}, \mathrm{Si}$) \cite{zhu2021topological}, using techniques such as the neutron scattering \cite{maleyev1995investigation, maleyev1998first, simonet2012magnetic, lee2013proposal} and the resonant inelastic x-ray scattering \cite{ko2011proposal, xiong2020resonant}. Additionally, Pyrochlore ferromagnets, exemplified by $X_2$V$_2$O$_7$ ($X = \mathrm{Lu}, \mathrm{Ho}$) and In$_2$Mn$_2$O$_7$ \cite{onose2010observation, ideue2012effect}, which are known to exhibit the magnon thermal Hall effect, are also considered potential candidate materials.

Given that SSC plays an essential role in Hall-type transport, our proposal suggests the possibility of such transport in collinear spin systems beyond the thermal Hall effect caused by topological magnons. One potential scenario is the topological Hall effect of conduction electrons driven by the magnon-induced SSC in collinear spin systems. Previous studies have shown that SSC in non-coplanar spin textures can induce the topological Hall effect of conduction electrons \cite{taguchi2001spin, tatara2002chirality, machida2010time}. However, based on our proposal, such a Hall effect could also arise in collinear spin textures. Another possibility is the phonon thermal Hall effect mediated by skew scattering due to the magnon-induced SSC. Recent studies have demonstrated the potential for the phonon thermal Hall effect arising from SSC in non-coplanar spin structures \cite{oh2024phonon}. By the same mechanism, we anticipate that the phonon thermal Hall effect driven by the magnon-induced SSC could also be present in collinear spin structures.

Indeed, recent experiments have suggested that the magnon-induced vector spin chirality contributes to the origin of nonreciprocal longitudinal transport in electronic systems \cite{nakamura2024nonreciprocal}. This finding raises the expectation that the magnon-induced SSC could also serve as a potential origin of transverse transport of elementary excitations such as electrons and phonons.

\medskip

\begin{acknowledgments}
This work was supported by Brain Pool Plus Program through the National Research Foundation of Korea funded by the Ministry of Science and ICT (2020H1D3A2A03099291). N.E. was supported by Forefront Physics and Mathematics Program to Drive Transformation (FoPM), a World-leading Innovative Graduate Study (WINGS) Program, the University of Tokyo and JSR Fellowship, the University of Tokyo. G. G. was supported by the National Research Foundation of Korea (NRF-2022R1C1C2006578).
\end{acknowledgments}

\medskip

\appendix

\begin{widetext}

\section{Derivation of Eqs. (\ref{Eq:SSC_eq_analytical_Kagome}) and (\ref{Eq:SSC_eq_analytical_Honeycomb})}
\subsection{Kagome ferromagnet case (Eq. (\ref{Eq:SSC_eq_analytical_Kagome}))} \label{Appendix:Derivation_Kagome}
Here we derive the analytical expression of the magnon-induced SSC (\ref{Eq:SSC_eq_analytical_Kagome}) by focucing only on the contribution near the band bottom. The lowest eigenenergy and eigenstate of the Hamiltonian (\ref{Eq:hk_Kagome}) around $\Gamma$ are given by \cite{katsura2010theory}
\begin{equation}\label{Eq:Kagome_E_approx}
    \epsilon_{1,\bm{k}} \simeq JSk^2 + 2KS + g\mu_B H,
\end{equation}
and
\begin{equation}\label{Eq:Kagome_vec_approx}
    \ket{u_{1,\bm{k}}} \simeq \frac{1}{W_{\bm{k}}} \begin{pmatrix}
        \lambda^{2}_{\bm{k}} - \cos^2 {k_2} \\
        \lambda^{2}_{\bm{k}} \cos{k_1} + \cos{k_2}\cos{k_3} \\
        \lambda^{2}_{\bm{k}} \cos{k_3} + \cos{k_1}\cos{k_2}
    \end{pmatrix}
    - \frac{i\phi}{W_{\bm{k}}} \begin{pmatrix}
        0 \\
        \lambda_{\bm{k}} \cos{k_1} - 2\cos{k_2}\cos{k_3}\\
        -\lambda_{\bm{k}} \cos{k_3} + 2\cos{k_1}\cos{k_2}
    \end{pmatrix},
\end{equation}
respectively, where $\lambda_{\bm{k}} \simeq 2(1-\frac{1}{6}\sum_{j=1}^{3} k^2_{j})$, and $W_{\bm{k}}$ is a normalization factor.
By combining Eqs. (\ref{Eq:chi_k_Kagome}) and (\ref{Eq:Kagome_vec_approx}), we obtain the equilibrium SSC profiles associated with the lowest band as

\begin{equation}\label{Eq:SSC_analytical_Kagome}
    \chi_{1,\bm{k}} \simeq -\frac{D_z}{9\sqrt{J^2 + D^{2}_{z}}S} (k^2_x + k^2_y)^2.
\end{equation}
Here, we only consider the lowest even order in $k_x$ and $k_y$ since the odd order terms do not contribute to the momentum space integration in Eq. (\ref{Eq:Magnon_induced_SSC_def_Kagome}). By substituting Eqs. (\ref{Eq:Kagome_E_approx}) and (\ref{Eq:SSC_analytical_Kagome}) into Eq. (\ref{Eq:Magnon_induced_SSC_def_Kagome}), we can calculate an approximate analytical expression for the magnon-induced SSC as

\begin{align}\notag
   \chi_{\mathrm{eq}} &\simeq \frac{1}{N_{\mathrm{cell}}}\frac{-D_z}{9\sqrt{J^2 + D^2_z}S} \sum_{\bm{k}}  \frac{(k^2_x + k^2_y)^2}{e^{\beta(2KS+g\mu_B H+JSk^2)}-1}\\ \notag
   &= \frac{-\sqrt{3}D_z}{18\pi^2\sqrt{J^2 + D^2_z}S} \sum_{\bm{k}} \frac{\pi}{N_x} \frac{2\pi}{\sqrt{3}N_y} \frac{(k^2_x + k^2_y)^2}{e^{\beta(2KS+g\mu_B H+JSk^2)}-1}\\ \notag
   &\simeq \frac{-\sqrt{3}D_z}{18\pi^2\sqrt{J^2 + D^2_z}S} \int_{\mathrm{BZ}} d^2 k \frac{(k^2_x + k^2_y)^2}{e^{\beta(2KS+g\mu_B H+JSk^2)}-1}\\ \notag
   &= -\frac{\sqrt{3} D_z}{9\pi\sqrt{J^2 + D^2_z}S} \int_{0}^{\infty} dk \frac{k^5}{e^{\beta(2KS+g\mu_B H+JSk^2)}-1}\\ \notag
   &= -\frac{\sqrt{3} D_z}{18\pi\sqrt{J^2 + D^2_z}S} \int_{0}^{\infty} dt \frac{1}{(\beta JS)^3} \frac{t^2}{e^{\beta(2KS+g\mu_B H)+t}-1}\\ \label{Eq:SSC_eq_analytical_derivation_Kagome}
   &= -\frac{\sqrt{3} D_z}{9\pi\sqrt{J^2 + D^2_z}S} \frac{\mathrm{Li}_3 (e^{-\beta(2KS + g\mu_B H)})}{(\beta JS)^3} \simeq -\frac{\sqrt{3} D_z}{9\pi\sqrt{J^2 + D^2_z}S} \frac{e^{-\beta(2KS + g\mu_B H)}}{(\beta JS)^3},
\end{align}
In going from the first and second lines to the third, we use $N_{\mathrm{cell}} = N_x N_y$, $dk_x \simeq \frac{\pi}{N_x}$, and $dk_y \simeq \frac{2\pi}{\sqrt{3}N_y}$ (we consider the deformed Brilluin zone: $[0,\pi)\times [-\frac{\pi}{\sqrt{3}},\frac{\pi}{\sqrt{3}})$). From the third to the fourth line, the integration over the Brillouin zone is replaced with that over all $\bm{k}$. Finally, in going from the fifth to the sixth line, we use the identity $\int_{0}^{\infty} dt \frac{t^s}{e^{t-\mu}-1} = \Gamma(s+1)\mathrm{Li}_{1+s}(e^{\mu})$, where $\Gamma(x)$ is the Euler gamma function and $\mathrm{Li}_n(z)$ is the polylogarithm function. Additionally, we apply the relation $\Gamma(3) = 2$ and the approximation $\mathrm{Li}_n(z) \simeq z$ for $|z| \ll 1$. 


\subsection{Honeycomb ferromagnet case (Eq. (\ref{Eq:SSC_eq_analytical_Honeycomb}))} \label{Appendix:Derivation_Honeycomb}
The lowest eigenenergy and eigenstate of the Hamiltonian (\ref{Eq:Magnon_Hamiltonian_Honeycomb}) are given by
\begin{equation}\label{Eq:Honeycomb_E}
    E_1 (\bm{k}) = \tilde{\epsilon} - |\bm{R}|
\end{equation}
and
\begin{equation}\label{Eq:Honeycomb_vec}
    \ket{u_1 (\bm{k})} = \frac{1}{\sqrt{2R(R+R_z)}} \begin{pmatrix}
        -R_x + iR_y\\
        R_z + R
    \end{pmatrix},
\end{equation}
respectively, where $\bm{R} = (R_x, R_y, R_z)$ with $R_x = -JS\sum_{j=1}^{3} \cos(\bm{k}\cdot \bm{c}_j)$, $R_y = JS\sum_{j=1}^{3} \sin(\bm{k}\cdot \bm{c}_j)$, and $R_z = -2D_z S\sum_{j=1}^{3} \sin(\bm{k}\cdot \bm{d}_j)$. Combining Eqs. (\ref{Eq:chi_k_Honeycomb_A}) and (\ref{Eq:Honeycomb_vec}) and expanding near the $\Gamma$ point yields the equilibrium SSC profiles for the lowest band at the lowest even order in $k_x$ and $k_y$ as
\begin{equation}\label{Eq:SSC_analytical_Honeycomb}
    \chi^{A}_{1,\bm{k}} \simeq \frac{9D_z}{32JS} k^2_x (k^2_x - 3k^2_y)^2.
\end{equation}
Inserting Eqs. (\ref{Eq:Honeycomb_E}) and (\ref{Eq:SSC_analytical_Honeycomb}) into Eq. (\ref{Eq:Magnon_induced_SSC_A_def_Honeycomb}) allows us to derive an approximate analytical expression for the magnon-induced SSC of the form 

\begin{align}\notag
   \chi^{A}_{\mathrm{eq}} &\simeq \frac{1}{N_{\mathrm{cell}}}\frac{9D_z}{32JS} \sum_{\bm{k}} \frac{k^2_x(k^2_x - 3k^2_y)^2}{e^{\beta(\frac{3}{4}JSk^2 + 2KS + g\mu_B H)}-1} \\ \notag
   &= \frac{27\sqrt{3}D_z}{256\pi^2 JS} \sum_{\bm{k}} \frac{2\pi}{\sqrt{3}N_x} \frac{4\pi}{3N_y} \frac{k^2_x(k^2_x - 3k^2_y)^2}{e^{\beta(\frac{3}{4}JSk^2 + 2KS + g\mu_B H)}-1} \\ \notag
   &\simeq \frac{27\sqrt{3}D_z}{256\pi^2 JS} \int_{\mathrm{BZ}} d^2 k \frac{k^2_x(k^2_x - 3k^2_y)^2}{e^{\beta(\frac{3}{4}JSk^2 + 2KS + g\mu_B H)}-1} \\ \notag
   &=\frac{27\sqrt{3}D_z}{256\pi^2 JS}\int_{0}^{\infty} k dk \int_{0}^{2\pi} d\theta \frac{k^6 (\cos^2 \theta (\cos^2 \theta - 3\sin^2 \theta)^2)}{e^{\beta(\frac{3}{4}JSk^2 + 2KS + g\mu_B H)}-1}\\
   \notag
   &= \frac{27\sqrt{3}D_z}{256\pi JS} \int_{0}^{\infty} dk \frac{k^7}{e^{\beta(\frac{3}{4}JSk^2 + 2KS + g\mu_B H)}-1} \\ \label{Eq:SSC_eq_analytical_derivation_Honeycomb}
   & \simeq \frac{\sqrt{3} D_z}{\pi JS} \frac{e^{-\beta(2KS+g\mu_B H)}}{(\beta JS)^4}.
\end{align}
To derive the third line from the first and second, we use $N_{\mathrm{cell}} = N_x N_y$, $dk_x \simeq \frac{2\pi}{\sqrt{3}N_x}$, and $dk_y \simeq \frac{4\pi}{3N_y}$ (we consider the deformed Brilluin zone: $[0,\frac{2\pi}{\sqrt{3}})\times [-\frac{2\pi}{3},\frac{2\pi}{3})$). To obtain the fifth line from the third and fourth, we extend the integration from the Brillouin zone to the entire $\bm{k}$ space and use the identity $\int_{0}^{2\pi} d\theta \cos^2 \theta (\cos^2 \theta - 3\sin^2 \theta)^2 = \pi$. Finally, in going from the fifth to the sixth line, we apply the identity $\int_{0}^{\infty} dt \frac{t^s}{e^{t-\mu}-1} = \Gamma(s+1)\mathrm{Li}_{1+s}(e^{\mu})$, the relation $\Gamma(4) = 6$, and the approximation $\mathrm{Li}_n(z) \simeq z$ for $|z| \ll 1$.


\end{widetext}

\bibliography{Ref_magnon_induced_SSC.bib}

\end{document}